# $^{36}$Cl concentrations from polar ice cores set new constraints on the Carrington event

F. Mekhaldi[1*], C. I. Paleari[1], A. M. Smith[2], A. Aldahan[3], J. Beer[4], M. Christl[5], C. Vockenhuber[5], H. Hayakawa[6], M. Curran[7], T. Erhardt[8,9], C. Plummer[7], K. Simon[2], K. Wilcken[2], M. Zheng[10] and R. Muscheler[11]

[1] *Department of Physical Geography and Bolin Centre for Climate Research, Stockholm University, Stockholm, Sweden*
[2] *Centre for Accelerator Science, Australian Nuclear Science and Technology Organization, Lucas Heights, NSW, Australia*
[3] *Geosciences Department, United Arab Emirates University, Al Ain, 15551, United Arab Emirates*
[4] *Eawag, Duebendorf, Switzerland*
[5] *ETH Zurich, Laboratory of Ion Beam Physics, Switzerland*
[6] *Institute for Space-Earth Environmental Research, Nagoya University, Nagoya, Japan*
[7] *Australian Antarctic Program Partnership, Institute for Marine & Antarctic Studies, University of Tasmania, Hobart, TAS, 7004, Australia*
[8] *Institute of Geosciences and Frankfurt Isotope and Element Research Center, Goethe University, Frankfurt am Main, Germany*
[9] *Climate and Environmental Physics, Physics Institute & Oeschger Centre for Climate Change Research, University of Bern, Sidlerstrasse 5, 3012 Bern, Switzerland*
[10] *School of Geographical Sciences, Fujian Normal University, Fuzhou, PR China*
[11] *Department of Geology – Quaternary sciences, Lund University, Lund, Sweden*

# Summary

The Carrington event of 1859 CE is considered as one of the largest geomagnetic storms of the observational era, and often used as a benchmark for a worst-case scenario. Yet, there exists no robust evidence of an associated solar energetic particle event of a significant magnitude, based on measurements of cosmogenic radionuclides $^{10}$Be and $^{14}$C from ice cores and tree rings, respectively. In this study, we present two $^{36}$Cl records from Greenland with 2-year and 4-year resolution from the EGRIP and NGRIP ice-core sites, together with semi-annual $^{10}$Be data from EGRIP, as well as annual $^{10}$Be and $^{36}$Cl concentrations from the Dome Summit Site, Law Dome, East Antarctica. We observe no significant $^{36}$Cl concentration increase around 1859 CE in the three records. This allows us to rule out an extreme solar energetic particle event hitting Earth associated with the Carrington event in terms of fluence above 30 MeV. Based on these ice core $^{36}$Cl measurements, we can suggest two scenarios: i) a soft SEP event with a maximum fluence above 30 MeV up to three times larger than any Space Age event or, ii) the possibility that there was no Earth-bound SEP event.

*Author for correspondence (florian.mekhaldi@natgeo.su.se).

# Main Text

### 1 Introduction

The Carrington flare occurred on September 1st, 1859 CE. It was the first solar flare to ever be directly observed, witnessed by astronomers R. Carrington and R. Hodgson while they were studying sunspots. Carrington described two patches of bright and white light emanating from a sunspot group [1–3] observed near the center of the solar disk. Following the solar flare, a disturbance of magnetic instruments at Kew (UK) was reported [4,5]. This solar eruption unleashed one of the greatest geomagnetic storms in the observational era, likely following the potential arrival of a coronal mass ejection (CME) [6,7]. As a consequence of the CME, the Carrington event caused disruptions of the telegraph system, in some cases allegedly giving the operators electric shock [8], and auroras were sighted in various parts of the world, further equatorward than normally expected, such as East Asia, Central America, South America, and the Pacific Islands [2,9]. In general, large solar flares and associated CMEs can accelerate solar energetic particles (SEPs), a subset of which may reach Earth and, in rare cases, produce ground level enhancements (GLEs) detected by neutron monitors.

The Carrington event is frequently cited as a representative example of an extreme or worst-case scenario pertaining to space weather impacts. In spite of its extreme effects, the detection of the Carrington event in proxy records has been somewhat challenging. The measurement of cosmogenic radionuclides in ice cores and tree rings is now a well-established technique to study solar storms pre-dating the observational era. To date, several extreme SEP events have been inferred from radionuclide measurements from tree rings and ice cores throughout the Holocene, such as in 774/5 CE and 993/4 CE [10–16], around 660 BCE [17], in 5259 BCE and 7176 BCE [18,19] and during the Last Glacial [20,21]. Cosmogenic radionuclides are produced by nuclear interactions between galactic cosmic rays (GCR), or SEPs emitted in connection with solar storms, with the atmosphere. Cosmogenic radionuclides such as $^{10}$Be and $^{36}$Cl are deposited onto ice sheets, while $^{14}$C is incorporated into the carbon cycle. Hence, the concentration of these isotopes can be retrieved and measured from ice cores and tree rings, respectively. The published $^{10}$Be data from the Dye 3 [22] and NGRIP [23] ice cores do not show any peak in association with the Carrington event. Similarly, no significant and ubiquitous increase was detected in Δ$^{14}$C records from tree rings [24], although an increase dated to 1859 CE in polar tree rings was recently reported [25]. Due to the lack of a significant increase in $^{10}$Be, Smart et al. [26] hypothesized that the event was likely not characterized by a hard spectrum, i.e. with a high proportion of high energy solar particles (>200 MeV) compared to lower energetic particles (~30 MeV). That is because events with a relatively low flux of high energy particles (i.e., soft events) do not contribute significantly to the production of cosmogenic $^{10}$Be [27–29]. For instance, the ground level enhancement no.05 in early 1956, characterized by a hard spectrum [30], is estimated to have caused an increase of 5.1% in the annual global $^{10}$Be production rate [28]. On the other hand, GLE no. 24 from 1972, characterized by a soft spectrum [31], is estimated to have caused an increase of only 1.2% in the annual global $^{10}$Be production rate [28]. The GLEs from 1972 occurred between the Apollo missions 16 and 17 and could have had severe consequences for the crews [32]. Moreover, there exists a softening of SEP event spectra towards the center of the Sun [33], from where the Carrington event originated, and soft events tend to induce a larger relative production rate increase in $^{36}$Cl compared to

$^{10}$Be and $^{14}$C [27,28]. In addition to cosmogenic radionuclides, several studies investigated the use of nitrate measurements from ice cores as a proxy for SEP events [34]. It was however found that nitrate peaks rather relate to other environmental phenomena such as biomass burning and that peaks tentatively linked to SEP events are not reproduced in various ice cores [35,36]. Furthermore, no nitrate peaks were observed around the 774/5 CE event [36,37].

In this study, we investigate new high-resolution $^{10}$Be and $^{36}$Cl records from Greenland and Antarctic ice cores around the Carrington event. In particular, we show new high-resolution $^{10}$Be and $^{36}$Cl data from the EGRIP S6 shallow ice core (East Greenland Ice core Project, 75°38′N, 36°00′W, 2704 m.a.s.l.), and new $^{36}$Cl measurements from the NGRIP 97-S2 shallow ice core (North Greenland Ice core Project, 75°6′N, 42°19′W, 2917 m.a.s.l.), together with previously published $^{10}$Be data from the same ice core [23]. We also present annual $^{10}$Be and $^{36}$Cl concentrations from the DSS0506 ice core in East Antarctica (Dome South Summit, 66°46'S, 112°48'E, 1375 m.a.s.l.).

## 2 Preparation and measurement of $^{10}$Be and $^{36}$Cl samples

The EGRIP S6 samples were prepared at the Department of Geology at Lund University (Sweden) following the procedure outlined by Paleari et al. [18]. After the addition of 0.15 mg of $^9$Be and 1 mg Cl carrier, the samples were filtered using ion exchange chromatography. Due to the lower concentrations of $^{36}$Cl, four EGRIP $^{10}$Be samples were combined into one $^{36}$Cl sample, containing a total of 4 mg Cl carrier. The NGRIP $^{36}$Cl samples were retrieved by combining four $^{10}$Be samples from the record presented in Berggren et al. [23] and were prepared at the Swiss Federal Institute of Aquatic Science and Technology (EAWAG) and at the Department of Earth Sciences at Uppsala University (Sweden). The $^{36}$Cl preparation of all samples was carried out following the procedure described in Delmas et al. [38].

The new $^{10}$Be and $^{36}$Cl samples presented here were measured with accelerator mass spectrometry (AMS) at the Laboratory of Ion beam Physics at ETH, Zurich, using the compact multi-isotope AMS system MILEA [39] and the 6 MV Tandem with the gas-filled magnet to suppress the isobar $^{36}$S [40], respectively. The measured $^{10}$Be/$^9$Be ratios were normalized to the ETH Zurich in house standards S2007N and S2010N, which were both calibrated relative to the ICN 01-5-1 standard ($^{10}$Be/$^9$Be = 2.709 × 10$^{-11}$ nominal) [39]. The measured $^{36}$Cl/Cl ratios were normalized to ETH Zurich in house standard K383/4N with a nominal value of (17.83 ± 0.35) x 10$^{-12}$.

The EGRIP S6 timescale is based on layer identification in the impurity record obtained from the continuous flow analysis (CFA) measurements collected at the Institute for Climate and Environmental Physics at the University of Bern. We estimate a timescale uncertainty of ± 1 year during the studied interval. The EGRIP record has an average resolution of 1.6 years for the $^{36}$Cl samples and 0.4 years for the $^{10}$Be samples. As detailed in Berggren et al. [23], the NGRIP 97-S2 timescale is constructed using annual layer counting in δ$^{18}$O data and it is synchronized to the GICC05 timescale [41], which is considered to not have any offset with the latest Greenland ice core chronology update (GICC21) at the time of the Carrington event [42]. The NGRIP 97-S2 timescale is considered to have an uncertainty of ± 1 year around the Carrington event [23].

As for the DSS0506 core, it was sampled at annual resolution, guided by the Australian Antarctic Division master DSS chronology. This Law Dome record was independently dated using annual layer counting of a suite of chemistry species with seasonally defined cycles [43], and the estimated uncertainty around the Carrington event is also ± 1 year. $^{10}$Be was extracted according to the procedure outlined in Pedro et al. [44], with the mobile phase retained for later $^{36}$Cl measurement. These samples were not filtered prior to preparation. Measurements of native Cl$^-$ in each sample was performed by ion chromatography (IC) using 5 mL aliquots. Chlorine carrier was then added to the samples, with the final Cl mass averaging 0.94 mg with native Cl$^-$ being approximately 14% of this. The solution was allowed to equilibrate before adding 1 mL of 50% HNO$_3$ and 0.5 mL of 10% AgNO$_3$. After 2 days precipitation in the dark the supernatant was discarded and the AgCl transferred and centrifuged after rinsing with milliQ water. The AgCl was dissolved in 5 mL of 22% NH$_3$ and 1 mL of saturated Ba(NO$_3$)$_2$ added before re-precipitating AgCl the next day with 5mL 69% HNO$_3$. After centrifuging and drying, the AgCl was pressed into AMS targets lined with AgBr. $^{10}$Be and $^{36}$Cl from DSS0506 were measured on the tandem 6 MV NEC accelerator (SIRIUS) at ANSTO's Centre for Accelerator Science (Australia), with the measurements corrected for procedural and machine background. For $^{10}$Be, KN-5-3 with nominal value of 6.320 x 10$^{-12}$ was used as primary standard, with KN[43]-5-4 (nominal value 2.851 x 10$^{-12}$) and KN-6-2 (nominal value 5.349 x 10$^{-13}$) used as secondary standards. For $^{36}$Cl, Z93-0005 with nominal value of 1.200 x 10$^{-12}$ (36/35=1.5837 x 10$^{-12}$, 36/37=4.9525 x 10$^{-12}$) was used as the primary standard.

## 3 Results

The $^{10}$Be and $^{36}$Cl concentration records from the EGRIP S6 and NGRIP 97-S2 ice cores in Greenland and from the DSS0506 core in Antarctica are shown in Fig. 1. The NGRIP $^{10}$Be record [23] was sampled at a resolution of 1 year and has an average concentration of 1.84 x 10$^4$ atoms/g ($\sigma$=16%) during the time span shown. The $^{36}$Cl record has a resolution of 4 years, and an average concentration of 0.22 x 10$^4$ atoms/g ($\sigma$=13%). The EGRIP $^{10}$Be record is characterized by an average concentration of 1.87 x 10$^4$ atoms/g ($\sigma$=25%). The EGRIP $^{36}$Cl record was sampled at an average resolution of 1.6 years, with an average concentration of 0.37 x 10$^4$ atoms/g ($\sigma$=18%). The DSS0506 record was sampled at a resolution of 1 year and has an average $^{10}$Be concentration of 6.86 × 10$^3$ atoms/g ($\sigma$=18%) and an average $^{36}$Cl concentration of 0.78 × 10$^3$ atoms/g ($\sigma$=18%). The average radionuclide concentrations are displayed as a dashed line in Fig. 1, whereas the standard deviation (+1$\sigma$) is shown as an orchid line for the $^{36}$Cl records.
As showcased in Fig. 1, all three records do not show any significant increase around 1859 CE, in agreement with previous findings from $^{10}$Be and $^{14}$C measurements [22–24]. Although the EGRIP S6 $^{10}$Be concentration record shows a peak in 1862 CE, we exclude such a large offset in the timescale and thus, do not consider it related to the Carrington event. This is corroborated by the fact that we do not find a synchronous $^{36}$Cl increase in either EGRIP S6, NGRIP 97-S2, or DSS0506. $^{10}$Be concentrations are characterized by high seasonal variability [44] with higher concentrations in Greenland and Antarctica during the summer season [45,46] and influenced by meteorological conditions [45,47,48]. Moreover, strong volcanic eruptions that release sulfate in the stratosphere may influence the deposition of $^{10}$Be onto the ice sheet [49,50]. It is thus worth mentioning that there was a large tropical eruption from Makian (Philippines) in 1861 seen in sulfate records from Greenland and Antarctica [51].

## 4 Discussion

### 4.1 Maximum fluence estimate

While the $^{36}$Cl measurements (Fig. 1) suggest the lack of an extreme SEP event, it remains possible that a significant event could have remained undetected in our ice core records as discussed by Mekhaldi et al. [28]. As first demonstrated by Webber et al. [27], intense SEP events can substantially increase the atmospheric production rate of $^{36}$Cl due to a resonance effect with target nuclei $^{40}$Ar for incident protons with energy around 30 MeV. More recently, the most efficient energy for the production (effective energy) of $^{36}$Cl by SEPs was refined to 60 MeV [29], based on new cosmogenic radionuclide production functions [52] coupled with revised GLE fluence spectra [53]. In contrast, the effective energies for $^{10}$Be and $^{14}$C production are 236 MeV and 234 MeV, respectively [29]. Consequently, the globally averaged production rate of $^{36}$Cl is particularly sensitive to soft-spectrum SEPs, which exhibit high fluence around 30-60 MeV and a rapid fall-off at higher energies. For instance, the SEP event of August 1972 (GLE no. 24) was calculated to have increased the global production rate of $^{36}$Cl by nearly 10 % relative to the baseline production by GCRs [28]. As mentioned above, the production rates of $^{10}$Be and $^{14}$C have been estimated to have increased by only 1 % for the same event. Considering the variance of the annual (DSS), biannual (EGRIP) and 4-year (NGRIP) $^{36}$Cl measurements shown in Fig. 1, it is visually evident that an annual 10 % increase would remain below the detection threshold set by the natural variability of the records (σ of 13%, 19% and 18% for NGRIP, EGRIP and DSS, respectively). Yet, the early August 1972 SEP event was exceptionally strong and remains the largest event on record in terms of fluence >30 MeV, with a Sun–Earth transit time of the related CME of about 14.6 hours [54]. Therefore, the absence of an observable $^{36}$Cl peak in any of the investigated ice cores does not necessarily imply that no significant SEPs accompanied the Carrington event.

      Based on the effective energy of each radionuclide [29], their respective GCR-induced baseline production rates [53], the depositional patterns of $^{10}$Be and $^{36}$Cl in polar ice [55,56], and the inherent noise in ice-core data, it has been suggested that an SEP event must have a fluence >30 MeV ($F_{30}$) of at least $5 \times 10^{10}$ protons/cm$^2$ to be detected in a single ice-core record above a 3σ threshold [28]. Here, we estimate the maximum possible fluence spectrum for a Carrington-related SEP event, given that no $^{36}$Cl enhancement is observed around 1859 CE at a >1σ level. We consider all recorded GLEs (https://www.nmdb.eu/nest/gle_list.php) for which reconstructed fluence spectra are available from Koldobskiy et al. (2021) and thus for which radionuclide production can be calculated [29], and which originated near the central meridian [33], similarly to the Carrington event within ±30° of solar longitude. To estimate the baseline $^{36}$Cl production, we use a solar modulation potential ϕ = 600 MV, representing the mean value for solar cycle no. 10 (i.e., the period encompassing the Carrington event) based on a new geomagnetic estimate of ϕ [57]. While the Carrington event occurred near the maximum of solar cycle no. 10, considering a solar cycle average value allows us to include effects such as radionuclide transport or meteorological and

post-depositional effects. The virtual axial dipole moment (VADM) is set to $8.5 \times 10^{22}$ Am² based on Nilsson et al. [58].

Fig. 2 shows scaled up fluence spectra (blue curves) of central meridian GLEs that would not yield a $^{36}$Cl concentration increase exceeding +20% of the baseline production by GCRs, which would otherwise be detectable. The inferred spectra are further complemented by boxplots at 30, 60, 100, 200, 500, and 1000 MeV illustrating the median values in red as well as the lower and upper quartiles (values shown in Table 2). The inferred spectra from Fig. 2 are also compared to the reconstructed fluence spectrum of the 774/5 CE event (black curve) [59], as well as spectra of modern GLEs (gray curves) [52]. The result includes atmospheric transport effects by applying a 54 % peak reduction corresponding to the average peak-to-integral production ratio derived from annual $^{10}$Be data for the 774/5 CE event [28]. We consider this peak reduction value to hold true for $^{36}$Cl as both radionuclides are expected to show similar deposition responses despite their different geochemical pathways [56]. It can be seen that a median $F_{30}$ of up to $1.75 \times 10^{10}$ protons/cm² would be consistent with the absence of a detectable $^{36}$Cl concentration peak. This value would correspond to a larger SEP event than the ones observed throughout the Space Age, but remain one order of magnitude smaller than the 774/5 CE event. This upper bound agrees with a $F_{30}$ estimate of ca. $10^{10}$ protons/cm² [60], inferred from an empirical relationship between soft X-ray (SXR) flux and $F_{30}$, though these values require revision as the source flare magnitude has been upscaled [3,61]. However, the fluence at higher energies (e.g., >1 GeV) would remain similar or lower than that of the largest GLEs on record, including GLE no. 05 (February 1956). Although a localized $\Delta^{14}$C increase has been reported in polar tree rings around 1859 CE [16], this signal is not reproduced in tree-ring records from other radiocarbon zones [24]. In the absence of a coherent global response and given the lack of coincident $^{36}$Cl enhancements, we do not interpret this feature as evidence for a Carrington-associated SEP event exceeding our fluence limits.

In summary, the $^{10}$Be and $^{36}$Cl concentration measurements from the NGRIP 97-S2, EGRIP S6 and DSS0506 ice cores (Fig. 1) constrain any potential Carrington-associated SEPs to have had, at most, a $F_{30}$ up to ca. 3 times that of the early August 1972 event. If we consider harder GLEs that occurred within 30° of the central meridian, such as GLEs no. 43 and 46 (Oct. - Nov. 1989), a $F_{30}$ in the order of $8 \times 10^9$ protons/cm² would be required which is only less than twice the $F_{30}$ of the August 1972 event. Such fluence >30 MeV would correspond to a severely high radiation dose rate for astronauts in interplanetary space [32], and likely cause widespread spacecraft disruptions, yet it would not constitute an extreme event in terms of GLE magnitude (Fig. 2), i.e., larger than Space Age GLEs.

### 4.2 On the likelihood of no associated SEP event

A scenario that cannot be ruled out is that the Earth was not reached by SEPs at the time of the Carrington event. Not all geomagnetic storms are associated with SEPs hitting the Earth. Gopalswamy et al. [62] mapped the solar locations of CMEs that caused intense geomagnetic storms and large SEP events between 1996 and 2016. They showed that geomagnetic storms are mainly associated with eruptive events originating close to the solar disk center. On the other hand, large SEP events are associated mostly with CMEs originating from the western limb of the Sun. That is because the western limb of the Sun is connected to the Earth through the heliomagnetic field lines [62]. Considering that the Carrington event originated near the center of the solar disk, it is possible that the SEPs did not hit the Earth, and therefore did not contribute to the production of cosmogenic radionuclides detected in ice cores and tree rings.

To assess such a likelihood, we have taken the 50 largest geomagnetic storms since 1957 CE, selected based on the hourly disturbance storm (Dst) index from the World datacenter for geomagnetism from Kyoto (https://wdc.kugi.kyoto-u.ac.jp/). From these, there remain 31 distinct events when merging geomagnetic storms occurring on consecutive days. Of these 31 largest geomagnetic storm events, ten occurred within ± 5 days of one or several GLEs (Table 1), although they may not arise from the same active region. This number remains the same whether we use ± 5-14 days. In other terms, about 32% of the largest geomagnetic storms of the Space Age were accompanied by a severe SEP event. It is therefore not surprising that the Carrington event may not have been associated with a GLE, or one that is large enough to be detected in $^{36}$Cl.

## 5 Conclusions

We have measured $^{36}$Cl concentrations at high resolution around the Carrington event of September 1859 CE for the first time, using data from three ice cores from Greenland (EGRIP S6 and NGRIP 97-S2) and Antarctica (DSS0506). Despite the enhanced production rate sensitivity of $^{36}$Cl relative to $^{10}$Be and $^{14}$C to SEPs, we do not detect any significant concentration increase around 1859 CE. This also holds true for seasonally resolved $^{10}$Be from the EGRIP shallow core. These results allow us to rule out an extreme SEP event, in terms of fluence >30 MeV, connected to the Carrington geomagnetic storm. We further assess a range maximum fluence spectra for a SEP event to not raise $^{36}$Cl concentrations above a detection threshold. In doing so, we estimate that a significant fluence >30 MeV would be possible, 3 times stronger than GLE no. 24 and one order of magnitude less than the 774/5 CE extreme SEP event. However, we cannot rule out that no Earth-bound SEPs were connected to the 1859 CE storm. In fact, we find that only about 32% of the largest geomagnetic storms of the Space Age were associated with a GLE. The $^{36}$Cl concentration records presented here thus set new constraints on the Carrington event, and contribute to a better understanding of the relationships between the different properties of space weather which have become increasingly important for our society.

## Acknowledgments

This research has been supported by the Royal Physiographic Society of Lund (grants to F.M.) and the Swedish Research Council (grant no. 2020-00420 to F.M and grants no. DNR2013-8421 and DNR2018-05469 to R.M.). R.M. further acknowledges funding from the European Union (ERC, PastSolarStorms, 101142677 to R.M.). Views and opinions expressed are however those of the author(s) only and do not necessarily reflect those of the European Union or the European Research Council. Neither the European Union nor the granting authority can be held responsible for them. H.H. thanks the financial support of JSPS Grant-in-Aids JP25K17436 and JP25H00635, the ISEE director's leadership fund for FYs 2021--2025, the Young Leader Cultivation (YLC) programme of Nagoya University, Tokai Pathways to Global Excellence (Nagoya University) of the Strategic Professional Development Program for Young Researchers (MEXT), the young researcher units for the advancement of new and undeveloped fields in Nagoya University Program for Research Enhancement, Transdisciplinary Network linking Space-Earth Environmental Science, History, and Archaeology Number (JPMXP1324134720) of MEXT Promotion of Development of a Joint Usage/Research System Project: Coalition of Universities for Research Excellence Program (CURE), and the NIHU Multidisciplinary Collaborative Research Projects NINJAL unit "Rediscovery of Citizen Science Culture in the

*Phil. Trans. R. Soc. A.*

Regions and Today". A.A. thanks the UPAR funding by the United Arab Emirates University. EGRIP is directed and organized by the Centre for Ice and Climate at the Niels Bohr Institute, University of Copenhagen. It is supported by funding agencies and institutions in Denmark (A. P. Møller Foundation, University of Copenhagen), USA (U.S. National Science Foundation, Office of Polar Programs), Germany (Alfred Wegener Institute, Helmholtz Centre for Polar and Marine Research), Japan (National Institute of Polar Research and Arctic Challenge for Sustainability), Norway (University of Bergen and Trond Mohn Foundation), Switzerland (Swiss National Science Foundation), France (French Polar Institute Paul-Emile Victor, Institute for Geosciences and Environmental research), Canada (University of Manitoba), and China (Chinese Academy of Sciences and Beijing Normal University). The NGRIP ice core project is directed and organized by the Ice and Climate Research Group at the Niels Bohr Institute, University of Copenhagen. It is supported by funding agencies in Denmark (SNF), Belgium (FNRS-CFB), France (IFRTP and INSU/CNRS), Germany (AWI), Iceland (RannIs), Japan (MEXT), Sweden (SPRS), Switzerland (SNF) and the United States of America (NSF). The DSS0506 core was drilled under Australian Antarctic Science Project 2384. Dr Mark Curran, Dr Andrew Moy, Meredith Nation and Chelsea Long of the Australian Antarctic Division are acknowledged for their assistance in providing the samples and the core chronology. The ANSTO Centre for Accelerator Science is funded by Australian Government National Collaborative Research Infrastructure Strategy.

# References

References will eventually be published in a numbered (Vancouver) format as shown below, but please format them as you wish for review. Please note that references to datasets must also be included in the reference list with DOIs where available.

# Tables

Table 1

|    | Date | Dst (nT) | GLE no. |
|----|------|----------|---------|
| 1  | 5/11/24 | -412 | 74 |
|    | 5/10/24 | -351 | 74 |
| 2  | 10/31/03 | -307 | 65, 66, 67 |
|    | 10/30/03 | -383 | 65, 66, 67 |
|    | 10/29/03 | -350 | 65, 66, 67 |
| 3  | 11/6/01 | -292 | 62 |
| 4  | 4/11/01 | -271 | 60 |
| 5  | 7/16/00 | -300 | 59 |
|    | 7/15/00 | -289 | 59 |
| 6  | 10/21/89 | -268 | 43, 44, 45 |
| 7  | 4/13/81 | -311 | 34 |
| 8  | 11/13/60 | -339 | 10, 11 |
| 9  | 4/30/60 | -325 | 8 |
| 10 | 7/16/59 | -283 | 7 |
|    | 7/15/59 | -429 | 7 |

Table 2

| Percentile/E | 30 | 60 | 100 | 200 | 500 | 1000 |
|---|---|---|---|---|---|---|
| Median | 1.75E+10 | 5.18E+09 | 1.98E+09 | 4.07E+08 | 3.33E+07 | 3.21E+06 |
| Q1 | 1.48E+10 | 4.62E+09 | 1.73E+09 | 2.88E+08 | 1.24E+07 | 8.75E+05 |
| Q3 | 2.23E+10 | 5.73E+09 | 2.27E+09 | 4.99E+08 | 4.72E+07 | 5.92E+06 |

# Figure and table captions

**Table 1**- List of geomagnetic storms, and their associated Dst index, within the top-50 that occurred between 1957-2024 CE associated with a GLE within ± 5 days.

**Table 2** – Integral fluences (>E) for the median, lower (Q1), and upper (Q3) quartiles of the scaled up GLE spectra representing the maximum fluence spectrum of a potential SEP event connected to the Carrington event shown in Fig. 2, and based on the data from Fig. 1. Values are in protons(>E)/cm$^2$.

**Figure 1**- $^{10}$Be (green) and $^{36}$Cl (orchid) concentrations from the EGRIP S6 ice core, in panel a, the NGRIP 97-S2 ice core in panel b, and the DSS0506 ice core from panel c. The dashed lines show the average radionuclide concentrations for each record, whereas the orchid horizonal lines shows the + 1 standard deviation of the $^{36}$Cl records. The year 1859 CE, when the Carrington event occurred, is highlighted in orange. 10Be from NGRIP 97-S2 (panel b) is from Berggren et al. [23].

**Figure 2**- Maximum fluence spectra estimates of a SEP event associated with the Carrington event (blue curves) considering the lack of a $^{36}$Cl concentration enhancement (Fig. 1), compared to the reconstructed fluence spectrum of the 774/5 CE event (black curve) [59] and of modern GLEs (gray curves) [52] where GLEs no. 24 and 56 are highlighted in orchid and green, respectively. The estimated fluence limits of the Carrington event are shown by boxplots (blue with median values in red) at 30, 60, 100, 200, 500, and 1000 MeV (Table 2).

# Figures

Figure 1

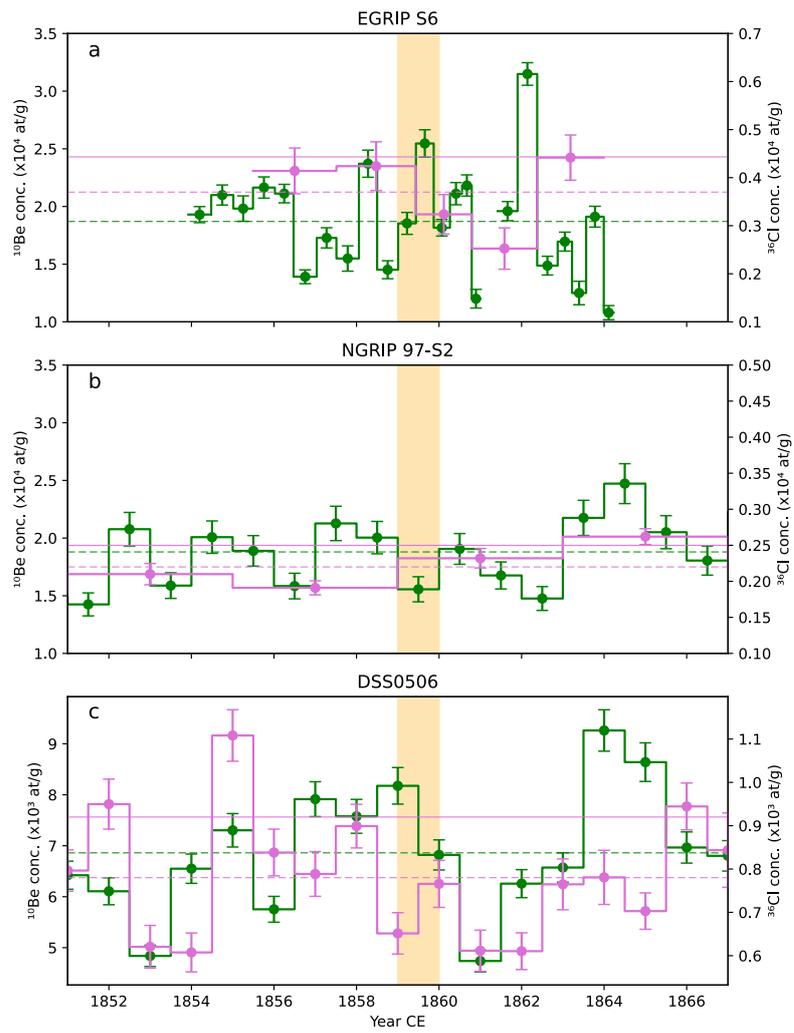

Figure 2

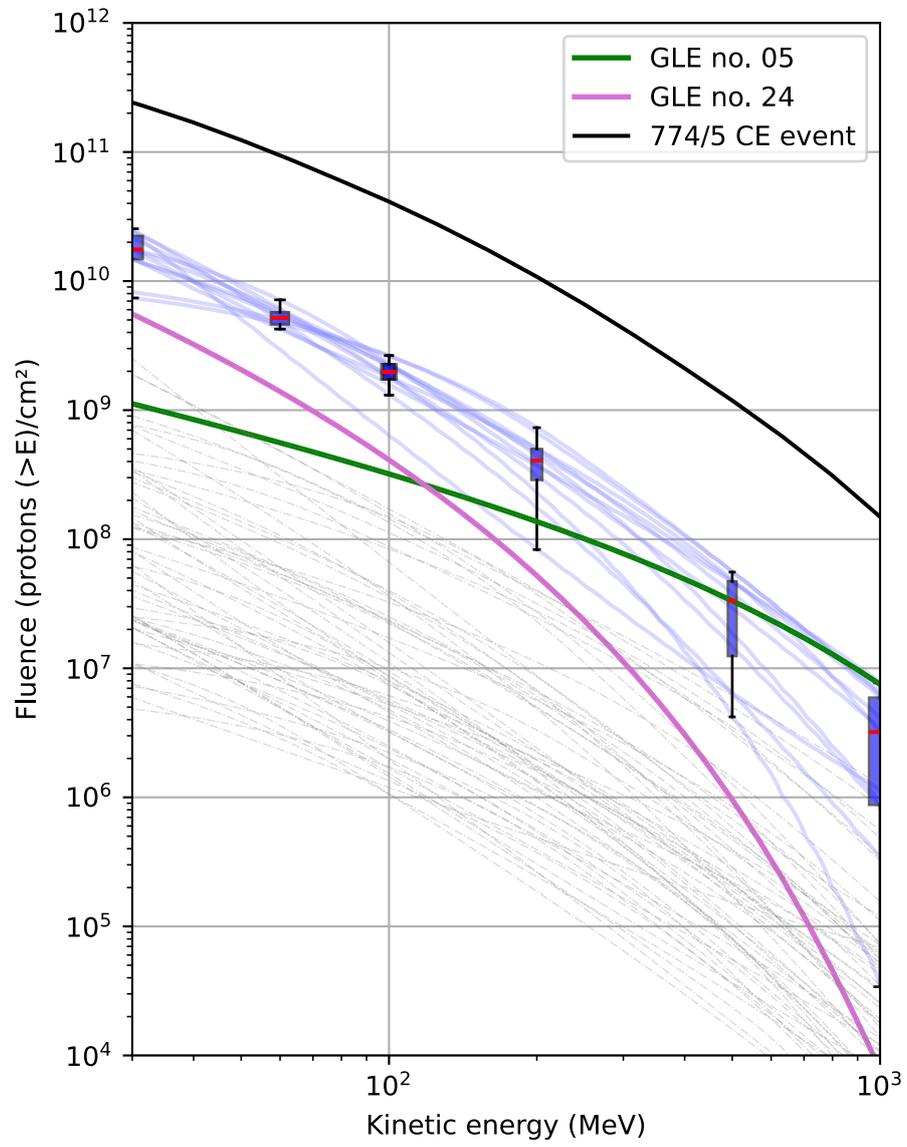